\documentclass{emulateapj}

\newcommand{\be}{  \begin{eqnarray} }
\newcommand{\ee}{  \end{eqnarray} }
\newcommand{\bd}{  \begin{displaymath} }
\newcommand{\ed}{  \end{displaymath} }
\newcommand{\msun}{ M_{\odot}}
\newcommand{\mdot}{\dot{M}}

\begin{document}
\title{Angular Momentum Transport in Accretion Disks and its Implications for
Spin Estimates in Black Hole Binaries}
\author{Chris Done\altaffilmark{1} and  Shane W. Davis\altaffilmark{2,3}}
\altaffiltext{1}{Department of Physics, University of Durham, South Road,
Durham, DH1 3LE, UK}
\altaffiltext{2}{School of Natural Sciences, Institute for Advanced Study, 
Einstein Drive, Princeton, NJ 08540}
\altaffiltext{3}{Chandra Fellow}

\begin{abstract}

The accretion flow in the disk dominated state of black hole binaries
has peak temperature and luminosity which vary together in such a way
as to indicate an approximately constant emitting area. The
association of this with the last stable orbit gives one of the few
ways to estimate spin when the mass of the black hole is
known. However, deriving this radius requires knowledge of how the
disk spectrum is modified by radiative transfer through the vertical
structure of the disk, as well as special and general relativistic
effects on the propagation of this radiation.  Here we investigate the
extent to which differences in vertical structure change the derived
disk spectra by calculating these for a range of different stress
prescriptions. We find that at a given mass accretion rate the spectra
are almost identical for accretion rates of $L/L_{Edd} \lesssim
0.1$. The spectra are remarkably similar even up to the highest
luminosities considered ($L/L_{Edd}\sim 0.6$) as long as the stresses
do not dissipate more than about 10 per cent of the gravitational
energy above the effective photosphere. This is exceeded only by
classic alpha disks with $\alpha\gtrsim 0.1$, but these models give
spectral variation which is incompatible with existing
data. Therefore, we conclude that disk spectral modelling can place
interesting constraints on angular momentum transport, but still
provide a robust estimate of the spin of the black hole.

\end{abstract}

\keywords{accretion, accretion disks --- black hole physics --- X-rays:binaries}

\section{Introduction}
 \label{intro}

Black hole spin is difficult to measure as it gives a strong signature
only very close to the event horizon. Currently the only way to probe
this is via luminous accretion flows, which light up the regions of
dramatically curved spacetime close to the event horizon. However, to
use these as a diagnostic of spin and/or to test of Einsteins gravity
in the strong-field limit requires knowledge of the velocity and
emission structure of this material. Currently, this is best
understood for an geometrically thin, optically thick accretion disk,
where the material rotates in approximately Keplerian orbits (except
at the inner boundary) and the gravitational energy thermalizes to
produce a quasi--blackbody spectrum at each radius in the disk
\citep{sas73}.

Black hole binaries (BHB) can indeed show spectra which look like this
simple `sum of blackbodies', though these are generally accompanied by
a nonthermal tail to higher energies whose origin is not well understood.
Nonetheless, where this tail carries only a minor fraction of the
bolometric power, then the disk model gives a fairly good description
of the spectral shape. More compellingly, a sequence of such disk
dominated spectra from the same object at different mass accretion
rates shows that the bolometric disk luminosity (spanning factors of
10--50) relates to the maximum temperature as $L_{\rm bol} \propto
T_{\rm max}^4$ \citep{kme01}, or equivalently, that the disk has an
approximately constant inner radius \citep{emh91,ebi93}. This {\em
observation} is exactly the behavior predicted by General Relativity,
as the last stable orbit forms a fixed size scale for a geometrically
thin disk \citep[see e.g. the review by][ hereafter DGK07]{dgk07}.

However, these disk dominated spectra {\em are} surprising in the
context of the Shakura--Sunyaev disk models as they are seen at high
luminosities, typically $0.1-0.5$ $L_{\rm Edd}$ (e.g. DGK07).  The
classic prescription in which the viscous stress is directly
proportional to total (gas plus radiation, $P_{\rm tot}=P_{\rm
gas}+P_{\rm rad}$) pressure, is thermally and viscously unstable where
radiation pressure dominates, i.e. for luminosities above $\sim 0.05
L_{\rm Edd}$ for BHB \citep[e.g.][]{lig74,sas76}. Above this
luminosity, the inner regions of alpha disks exhibit limit cycles, in
direct conflict with the observed {\em stability} of the disk spectra
\citep[e.g.][]{now95}. Yet the alpha prescription does seem to be able
to reproduce the long timescale outburst/quiescence behavior of the
outer, gas pressure dominated region in the disk instability model
\citep[see e.g. the review by][]{las01}.

Thus it seems likely that the stress is approximately proportional to
gas pressure where gas pressure dominates, but that it scales less
sensitively with temperature than predicted by radiation pressure when
radiation pressure dominates. Ultimately, the question of the stress
scaling will be answered by numerical simulations of the
Magneto-Rotational Instability \citep[MRI,][]{bah91} which forms the
physical basis for ``effective viscosity''.  However, this is beyond
the scope of current codes as it requires coupled
radiative--magneto--hydrodynamic simulations in full general
relativity, covering a large radial extent but simultaneously
resolving the small scale height of the thin disk!. Until these become
available, {\em ad hoc} stress scalings are the only way to model the
structure of a radiative accretion disk.

The choice of stress prescription can affect the spectrum as well as
the stability of the disk as it determines the surface density, and
therefore, the overall optical depth of the disk. This emission
thermalizes to a true blackbody only if the disk is effectively
optically thick to {\em absorption} at all frequencies.  Free-free
(continuum) and bound--free (photo-electric edge) absorption opacity
both drop as a function of frequency while electron scattering is
constant, so the highest energy photons from each radii are less
likely to thermalize.  The spectrum then becomes a modified blackbody
whose frequency dependence can, in general, differ significantly from
Planckian.  However, the effects of Compton scattering tend to yield a
Wien-like tail and the hottest parts of the disk are generally well
modelled by a color-corrected (or diluted) blackbody, with effective
temperature which is a factor $f_{\rm col}$ (termed a
color-temperature correction) higher than for complete
thermalization. The full-disk spectrum is then a sum of modified black
bodies, but this can likewise be approximately described by a
color-temperature corrected blackbody disk spectrum
\citep{sat95}. This color-temperature correction depends on the
vertical temperature and density structure of the disk, both of which
can only be fully determined once the stress is specified.

Since the observations show to zeroth order that $L\propto T^4$, the
color-temperature correction must stay approximately constant (the
alternative, that it changes in just such a way as to mask any change
in innermost extent of the disk, violates Occam's razor). This used to
be controversial, with different models of alpha disks (with only H
and He) giving different results. \cite{sat95} found that $f_{\rm
col}$ remains relatively constant over a wide range in luminosity,
while \cite{mfr00} had constant $f_{\rm col}$ only at $>0.1L_{\rm
Edd}$, {\em increasing} below \cite[ hereafter GD04]{gad04}. Since it
is physically rather unlikely that absorption becomes less effective
at lower temperatures, the \cite{mfr00} result seems rather to be an
artifact of the assumed constant vertical density structure. However,
both these calculations are now superseded by the models of
\cite{dav05}, which includes fully non-LTE calculations for all
abundant elements together with the self--consistent radiative
transfer and disk structure. These indeed show that $f_{\rm col}$ is
approximately constant for an alpha disk, but only below temperatures
of $\sim 1$~keV. Above this, the disk becomes so ionized that even
photo-electric absorption opacity becomes negligible, so $f_{\rm col}$
increases markedly \citep{dav05,ddb06}.  A constant $f_{\rm col}$ can
be recovered even at these high temperatures by simply decreasing
$\alpha$ to $\sim 0.01$, but this is inconsistent with the $\alpha\sim
0.1$ required in order to match the rapid rise to outburst seen in
transient systems \citep[e.g.][]{kpl07}, and still gives a disk which
should undergo limit cycle oscillations when radiation pressure
becomes dominant.

Thus it seems that both spectra and stability of the observed disk
spectra in BHBs are inconsistent with an alpha stress prescription.
Here we use the methods of \cite{dav05} and \cite{dah06} to examine
two alternative models, where the stress is directly proportional to
gas pressure alone, hereafter called the beta disk, or to the
geometric mean of the gas and total pressure, hereafter called the
mean disk, \citep{sac81,sar84,tal84,lan89,mer03}.  When we fix
$\alpha=0.1$ to agree with requirements of the disk instability model,
we find that either of these give a better match to the spectral
constraints than an alpha prescription as the resulting disk is denser
and more optically thick. Thus these alternative stress scalings give
disks which are able to maintain a relatively constant $f_{\rm col}$
even at high temperatures, providing a better match to the observed
spectra as well as the stability properties.

While this gives important insight into the approximate form of the
MRI stresses, we show that {\em all} stress prescriptions considered
here (alpha, beta, and mean) give the {\em same} $f_{\rm col}$ below
$\sim 1$~keV, suggesting that disk spectra can be used to give a
robust estimate of the inner radius of the disk, hence observationally
constraining the black hole spin.

\begin{figure*}
\begin{center}
\hbox{
\includegraphics[width=0.5\textwidth]{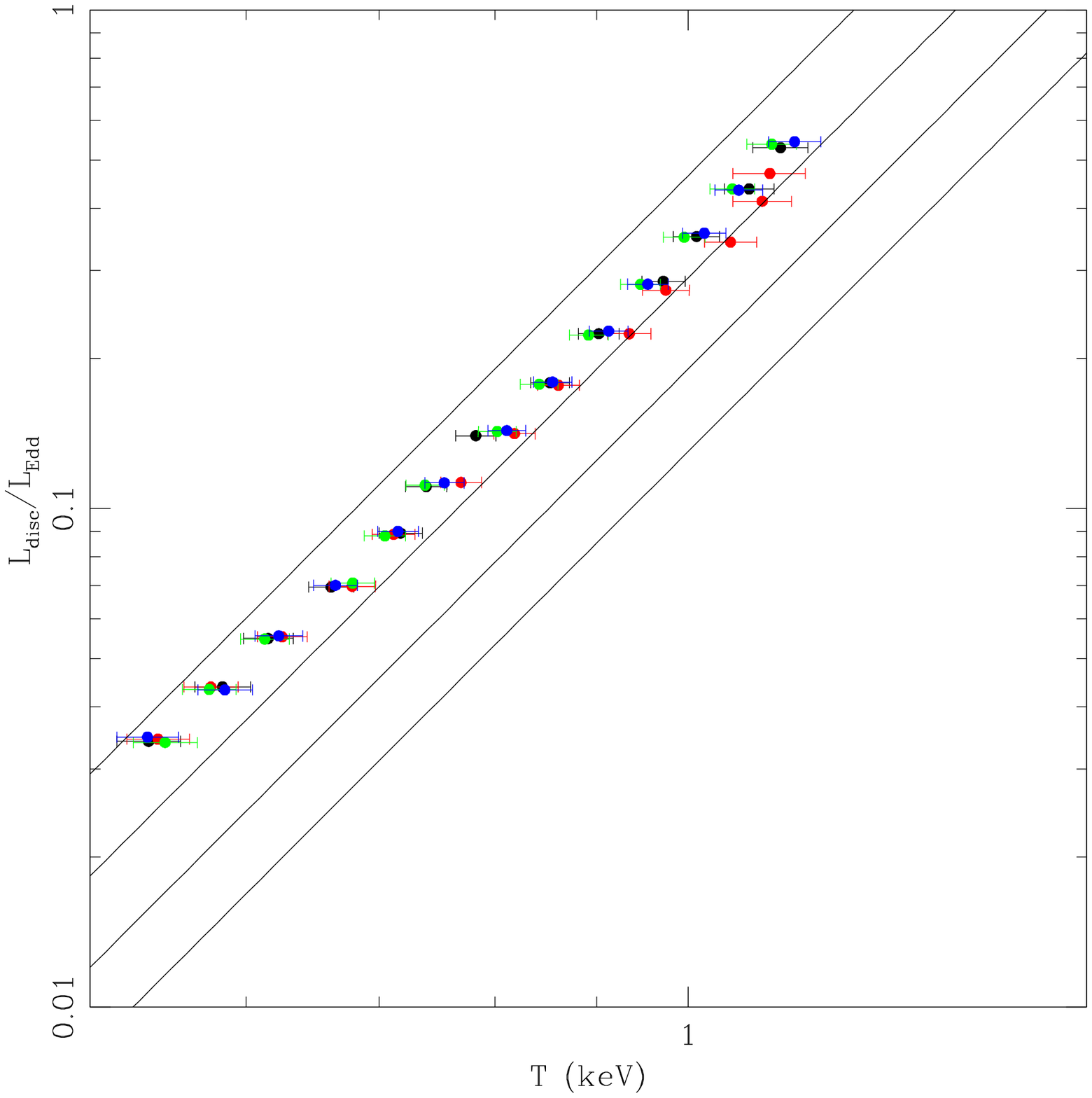}
\includegraphics[width=0.5\textwidth]{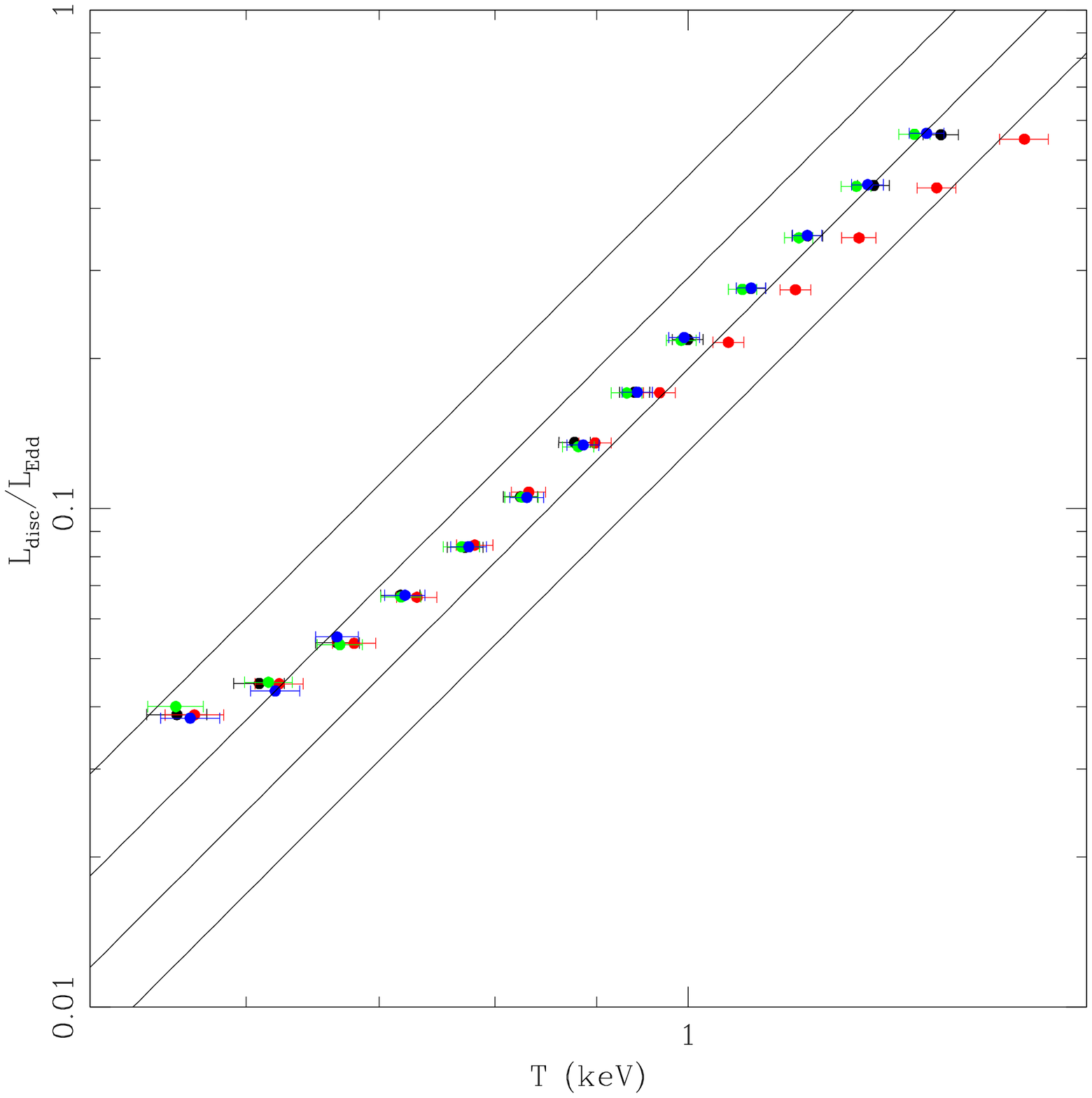}
}
\end{center}
\caption{Luminosity--Temperature relations derived from fits of 
{\tt bhspec} with {\tt diskbb} as described in \S\ref{fcol_0.5}.
The left panel corresponds to fits
using a CCD ({\it Swift} XRT) response while the right panel utilizes a
proportional counter ({\it RXTE} PCA) response.  We consider four different
stress prescriptions: alpha disks with $\alpha=0.01$ (black) and 
$\alpha=0.1$ (red), beta disks with $\alpha=0.1$ (blue), and mean
disks with $\alpha=0.1$ (green). These all overlap apart from the 
alpha disks with $\alpha=0.1$ at high luminosity, showing how robust
the spectra are to different stress prescriptions as long as the disk
remains optically thick. The solid black lines show
the $L-T$ relation expected from black hole with $a_*=0.5$ 
for a color-temperature correction of
1.6, 1.8, 2.0 and 2.2 from top to bottom, respectively.
\label{f:lt} }
\end{figure*}

\section{Alternative stress prescriptions}
\label{stress}

The magnetic nature of the viscosity has long been recognized
\citep{sas73,eal75,ich77,grv79}, motivating alternative stress
prescriptions \citep[e.g.][]{sac81}. The two most popular are where
stress is directly proportional to gas pressure alone \citep[beta
disks,][]{sar84} or to the geometric mean of the gas and total
pressure, $\sqrt{P_{\rm gas}P_{\rm tot}}$ \citep[mean
disks,][]{tal84,mer03}.  Beta disks are stable at all luminosities
below the Eddington limit (above which the equations break down as
winds and/or advection become important), while the mean disks show
limit cycles above $\sim 0.3 L_{\rm Edd}$ for stellar remnant black
holes \citep[e.g.][]{hkm91,man06}. The mean disk is thus inconsistent
with the observation that stable disk spectra exist up to at least
$\sim 0.7L_{\rm Edd}$ (DKG07), but the onset of the instability is
very sensitive to the exact scaling of the stress. For the more
general prescription which is proportional to $P_{\rm gas}^\mu P_{\rm
tot}^{1-\mu}$ then the observed stability limit requires $\mu\sim
0.56$ \citep{kfm98}, which is negligibly different to the mean disk
($\mu=0.5$) in terms of steady state disk structure.

We use the {\tt bhspec} code of \cite{dah06} to calculate the
self-consistent vertical structure and resulting spectra for three
stress prescriptions: alpha, beta, and mean disks.  We consider a
geometrically thin disk and break it up into a series of
logarithmically spaced annuli.  We then calculate the one-dimensional
vertical structure and radiative transfer in each annulus using the
TLUSTY stellar atmosphere code \citep{hal95}.  The total spectrum
is then integrated over the disk surface, accounting for the effects
of space-time curvature on the photon geodesics.  Three parameters
determine the structure of each individual annulus: the radiative flux
at the disk surface; the surface density, $\Sigma$; and a parameter
which determines the strength of the tidal gravity.  

Since mechanical
energy dissipates directly within the disk ``atmosphere'', the
radiative flux $F$ varies with height $z$ in the disk, requiring an
additional assumption to uniquely determine the structure.  Following
\cite{sas73}, we assume that the dissipation $d F/dz$ is locally
proportional to the density $\rho$. Using the definition of column
mass, $dm=-\rho dz$, this is equivalent to
\be
\frac{d F}{dm}=-\frac{2 \sigma T_{\rm eff}^4}{\Sigma},
\label{eq:dis}
\ee
where $T_{\rm eff}$ is the effective temperature of the annulus.  Note
that the right hand side terms, and therefore $dF/dm$ are independent
of height with this assumption.  We cannot use the stress prescription
to specify a local viscous-like dissipation as this is proportional to
some combination of radiation and gas pressure. These pressures are
largest at the disk midplane due to hydrostatic equilibrium, so the
dissipation is also largest there, and the resulting temperatures are
so high as to lead to the disk becoming Rayleigh-Taylor unstable. This
would lead to convection, presumably driving the disk structure
towards the marginally stable condition where the dissipation is
proportional to density.  Time and horizontal averages of shearing box
simulations \citep{tur04,hks06} indicate that a somewhat greater
fraction of the dissipation occurs at low column near the surface.

We use a vertically integrated disk model to determine the above
parameters for each annulus. The tidal gravity is specified by the
choice of spacetime and radius, while the radiative flux at the
surface is independent of stress prescription for a thin disk
\citep[see e.g.][]{sas73}.  Therefore, we only require a stress
prescription to determine the surface density in this model.  (Note,
however, that the theoretical arguments which motivate the stress
prescription may also require modifications to the dissipation
profile, a point we will discuss further in \S\ref{discus}.)  The
surface density may be determined by enforcing angular momentum
conservation
\be 
\int \tau_{r\phi} dz = 
\frac{\mdot \Omega}{2 \pi}\frac{D(r)}{A(r)}, 
\label{eq:coa} 
\ee
 where $\tau_{r\phi}$ is the accretion stress, $\mdot$ is the accretion
rate, and $\Omega$ is the Keplerian frequency. The functions $A(r)$
and $D(r)$ (both $\to 1$ for radii much larger than the last stable
orbit) defined in \cite{rah95} incorporate the no--torque (stress
free) inner boundary condition and parameterize the deviation of
general relativistic disk structure from the Newtonian limit.

\begin{figure*}
\begin{center}
\hbox{
\includegraphics[width=0.5\textwidth]{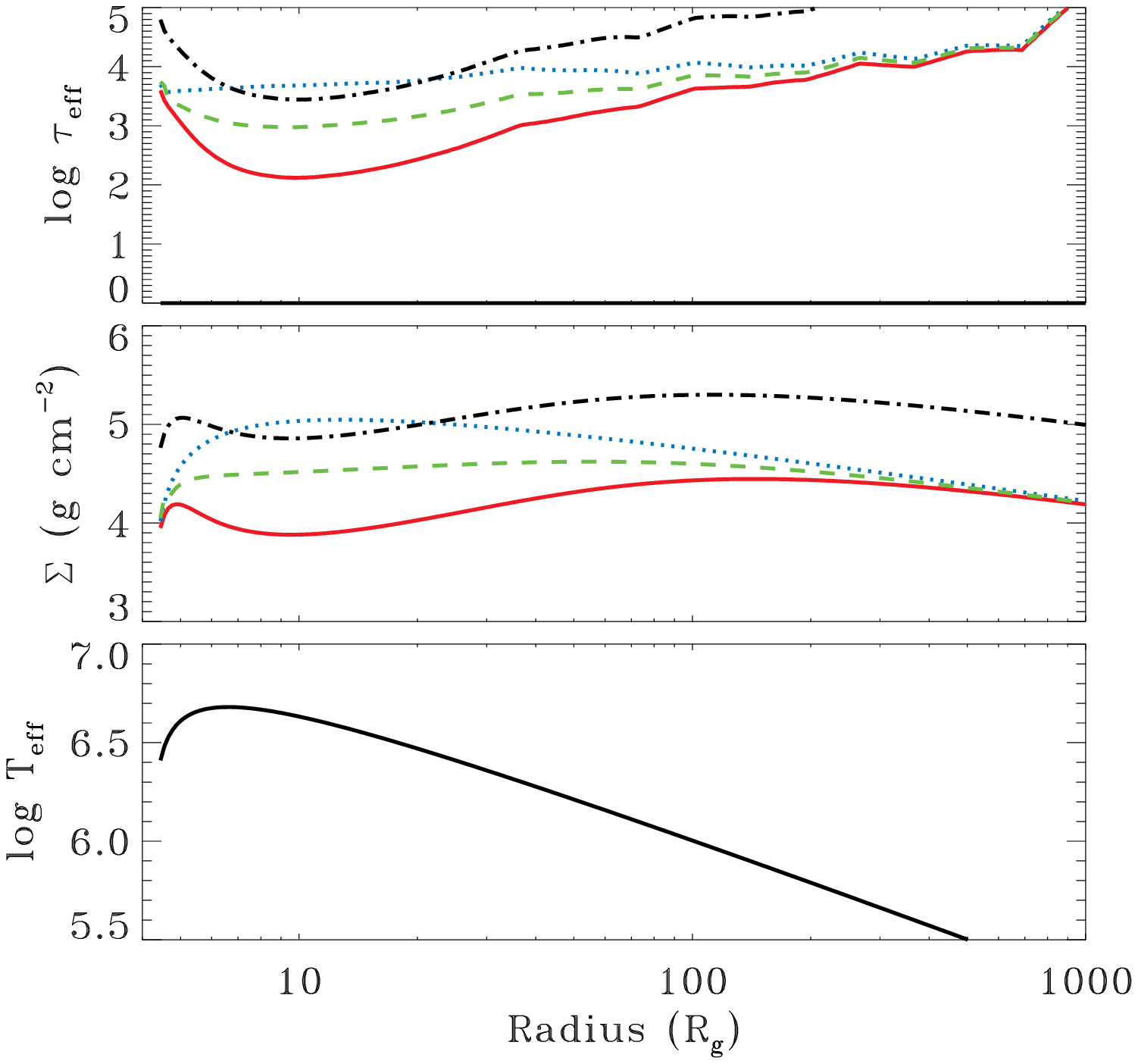}
\includegraphics[width=0.5\textwidth]{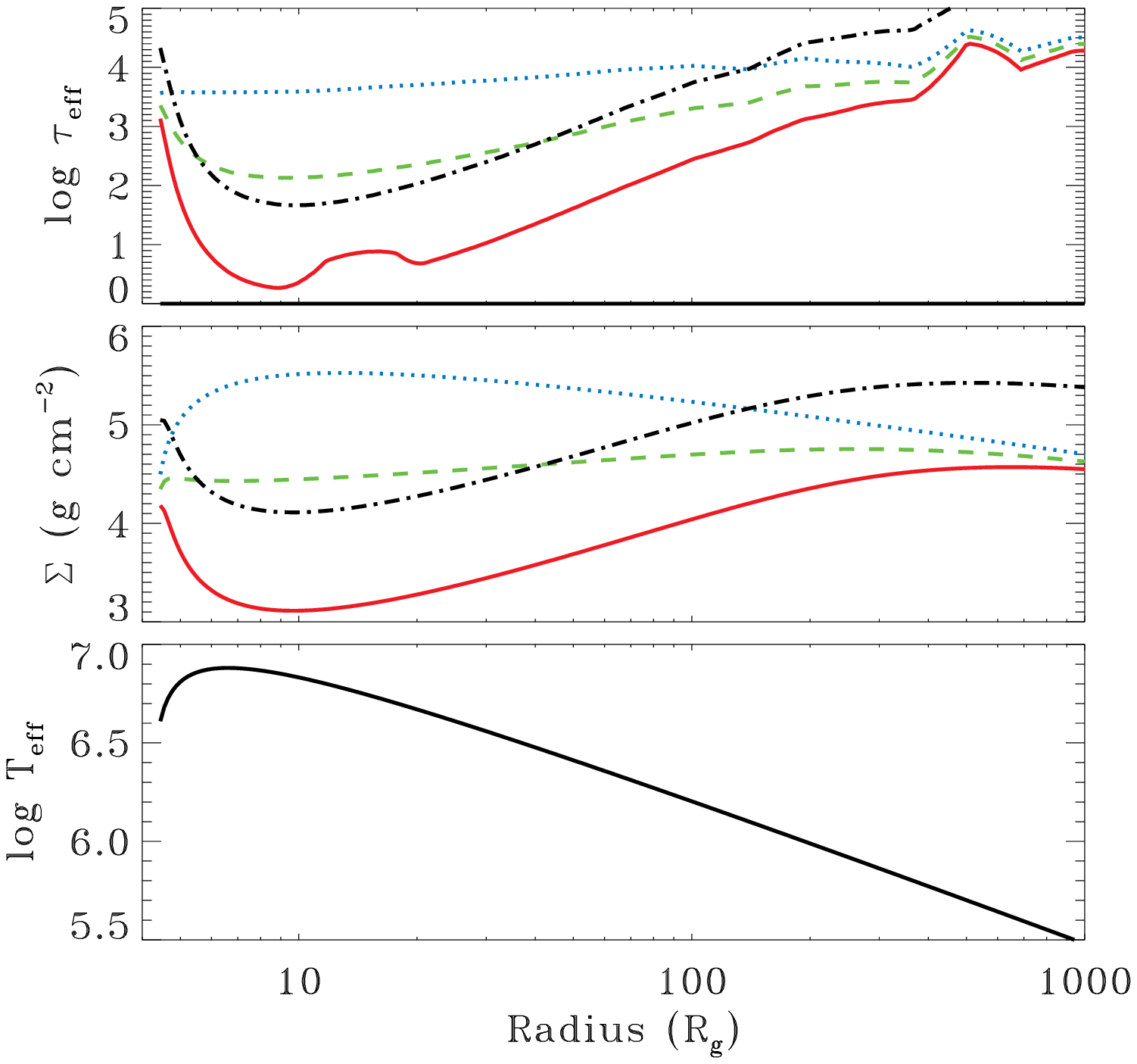}
}
\end{center}
\caption{The upper, middle and lower panels show the effective optical
depth to the disk midplane, total surface density and effective
temperature as a function of radius, respectively. We compare alpha
disks with $\alpha=0.01$ (black, dot-dashed) and $\alpha=0.1$ (red,
solid), beta disks
with $\alpha=0.1$ (blue, dotted), and mean disks with $\alpha=0.1$
(green, dashed). $T_{\rm eff}$ is independent of stress prescription.
(a) shows these disk parameters for $\log l=-1$ while (b) has $\log
l=-0.2$.\label{f:l-1}}
\end{figure*}

The integral in eq. (\ref{eq:coa}) presents a computational difficulty
since $\tau_{r\phi}$ depends on local variable such as temperature and
density for the stress prescriptions considered here.  Therefore, the
exact computation requires the full vertical structure which, in turn,
requires a choice of $\Sigma$.  Therefore, an exact solution can only
be computed via iteration provided an initial guess for $\Sigma$
\citep[see e.g.][]{dav05}.  Given the approximations inherent in our
stress prescriptions, we instead solve for $\Sigma$ using moderately
crude vertical averages to replace $\int\tau_{r\phi} dz$ with $2
H\langle\tau_{r\phi}\rangle$.  Here $H$ is an approximate vertical
scale height for the disk determined via the equation of hydrostatic
equilibrium
\be
\frac{\partial P_{\rm tot}}{\partial z}=\rho \Omega^2 z \frac{C(r)}{B(r)},
\label{eq:he}
\ee
where $B(r)$ and $C(r)$ are relativistic factors (again both $\to 1$
for radii much larger than the last stable orbit) defined in
\cite{rah95}.  This is solved by substituting $z \rightarrow H$ and
$\partial/\partial z \rightarrow H^{-1}$.  We then specify $\langle
P_{\rm rad} \rangle$ and $\langle P_{\rm gas} \rangle$ in terms of
$\Sigma$, yielding two equations (\ref{eq:coa}, \ref{eq:he}) for two
unknowns ($\Sigma$, $H$).  The resulting $\Sigma$ evolves smoothly
between the gas and radiation pressure dominated limits.  Previous
work \citep{dav05} has shown that the resulting estimates for $\Sigma$
typically agree with a more precise iterative method to $\lesssim
30\%$.

Our results also agree with the gas and radiation pressure dominated limits
which have been worked out in detail by previous authors
\citep[e.g.][]{sas73,sar84,lan89,mer03}. We give the asymptotic 
behavior at large radii 
(away from the inner boundary condition) below as this is useful in 
understanding the subsequent plots. For the gas pressure
dominated limit all three stress prescriptions give identical
results where 
\be
\Sigma_{\rm gas} \propto m^{1/5} l^{3/5} r^{-3/5} \alpha^{-4/5} 
\label{eq:siggas}
\ee
where $l\equiv L/L_{\rm Edd}$ is the Eddington ratio, $r \equiv R/R_g$
is radius in unit of the gravitational radius, and $m \equiv M/\msun$
is the black hole mass.  
For the radiation pressure dominated limit the surface density is sensitive to the
stress prescription, and for our choices gives
\be
\Sigma_{\rm rad, \alpha} & \propto & l^{-1} r^{3/2} \alpha^{-1}, \nonumber \\ 
\Sigma_{\rm rad, \beta} & \propto & m^{1/5} l^{3/5} r^{-3/5} \alpha^{-4/5}, \nonumber \\
\Sigma_{\rm rad, mean} & \propto & m^{1/9} l^{-1/9} r^{1/3} \alpha^{-8/9}.
\label{eq:sigrad}
\ee

Since the BHBs considered here cover only a small range in mass, the
weak dependence of $\Sigma$ on $m$ can be ignored.  In what follows,
the Eddington ratio dependence will be most important.  For the alpha
and mean disks $\Sigma$ decreases as $l$ increases, but with a much
weaker dependence for the mean disk.  In contrast, $\Sigma$ increases
with $l$ in the beta disk model.  The susceptibility of the stress
prescriptions to viscous instability follows directly from these
relations.  The condition for stability is that $d \Sigma/d \mdot > 0$
\citep{lae74}.  Since $\mdot \propto l$, we see that only the beta
disk is truly stable in the limit where radiation pressure dominates
entirely, though the mean disk is also very close to the stability
condition.

\begin{figure*}
\begin{center}
\hbox{
\includegraphics[width=0.5\textwidth]{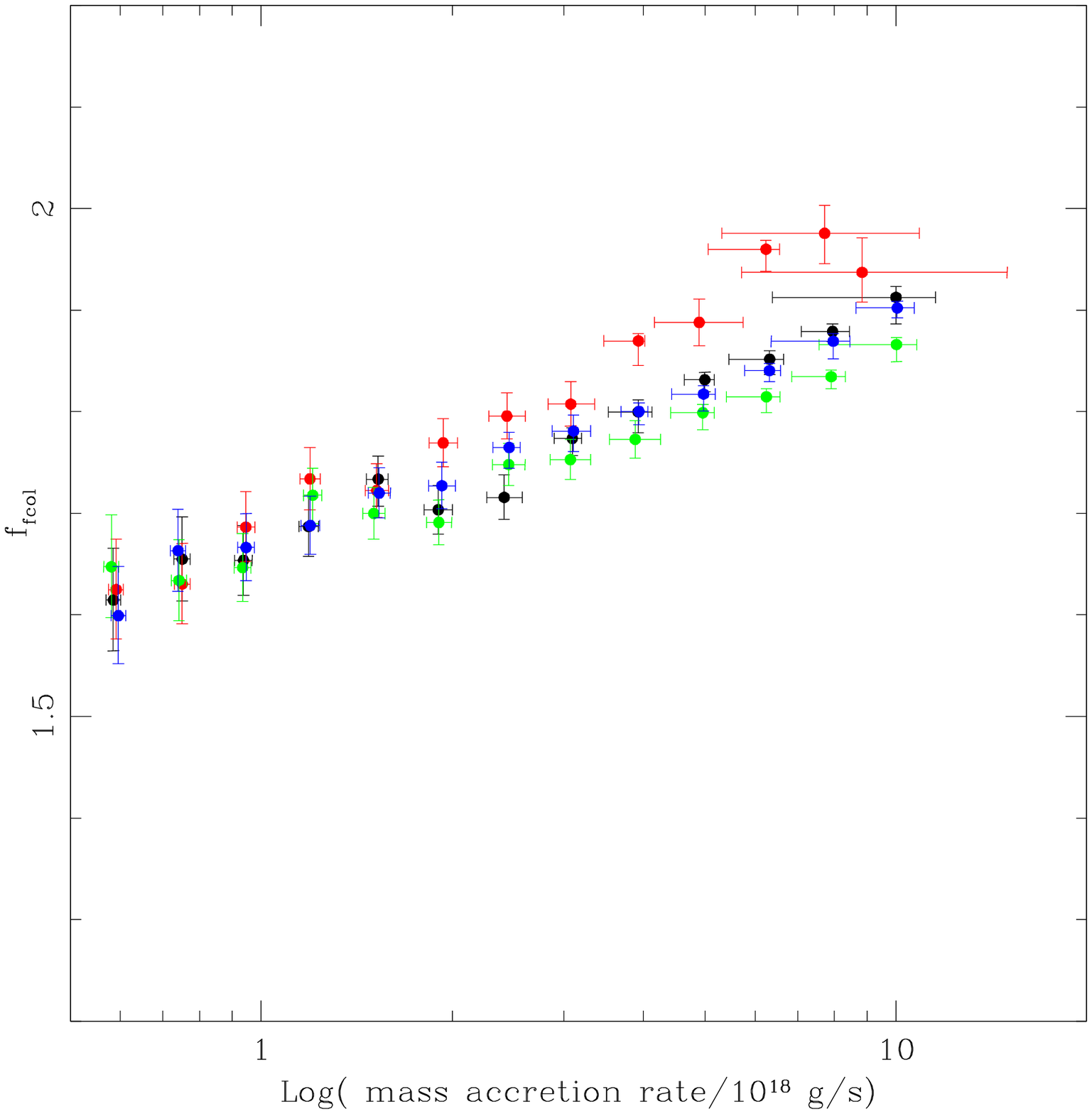}
\includegraphics[width=0.5\textwidth]{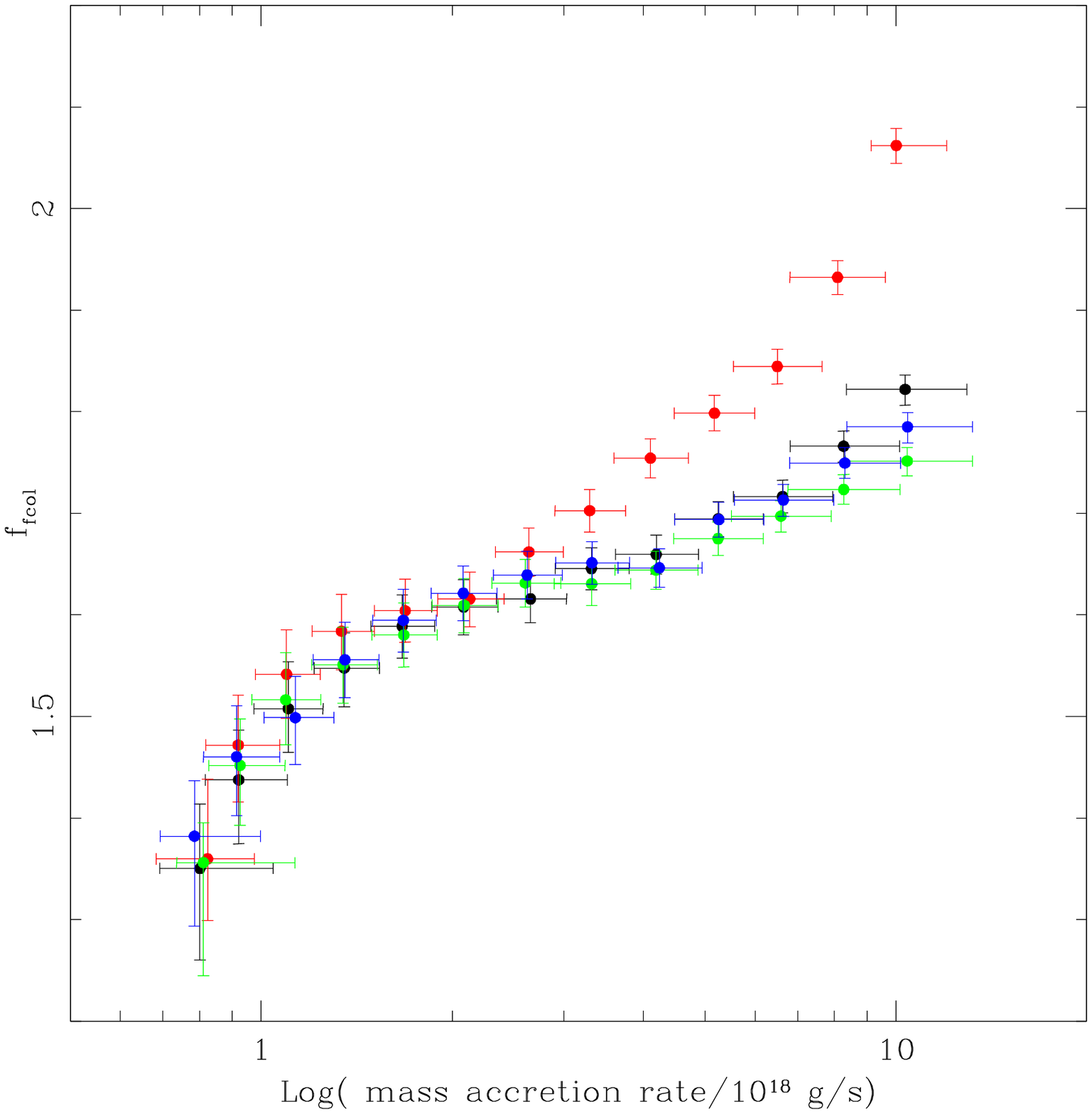}
}
\end{center}
\caption{Color-temperature corrections from fits of {\tt bhspec} with
{\tt kerrbb} as described in \S\ref{fcol_0.5}.  The left panel
corresponds to fits using a CCD ({\it Swift} XRT) response while the
right panel is for a proportional counter ({\it RXTE} PCA) response.
We consider four different stress prescriptions: alpha disks with
$\alpha=0.01$ (black) and $\alpha=0.1$ (red), beta disks with
$\alpha=0.1$ (blue), and mean disks with $\alpha=0.1$
(green).\label{f:fcol1}. Comparison with Fig 1 shows that the results
for the color-temperature correction are not the same as for {\tt
diskbb}, where the CCD bandpass gives an almost constant 
$f_{col}=1.65-1.75$ as opposed to the change from 1.6-1.9 seen here. 
The proportional counter likewise has $f_{col}=1.7-2.0$ with
{\tt diskbb} compared to $1.4-1.8$ here.}
\end{figure*}

\section{Luminosity-Temperature plots and the color-temperature
  correction}
\label{fcol_0.5}

We now use this {\tt bhspec} code \citep[ see also \S \ref{stress}]{dah06}
to investigate the
spectra produced by these different stress prescriptions for a black
hole of 10 $\msun$, inclined at 60$^\circ$, with spin of $a_*=0.5$. We
use the {\sc XSPEC} spectral fitting package to convolve the models with  
the detector response from current X-ray instruments in order
to simulate what would be observed. We do this for 
accretion rates equally spaced in $\log l$ from
-1.4 to -0.2 through both the {\it RXTE} PCA (3--20~keV) and {\it
Swift} XRT (0.1--10~keV) responses. These two instruments 
are broadly characteristic
of any proportional counter (e.g. GINGA, ASTROSAT) and CCD
(e.g. Chandra, XMM-Newton, Suzaku XIS) responses, respectively.
We fix the source distance at 8.5~kpc (Galactic center), and 
assume a Galactic absorbing column of $N_H=1.2\times 10^{21}$
cm$^{-2}$ (approximately the lowest possible at this distance). 
We note that for $\log l =-0.6$, this gives a very good
estimate of the disk dominated spectrum of XTE J1817-330 at the peak
of its outburst \citep[spectrum 001:][ hereafter GDP08]{ryk07,gdp08}.

We also include a small Comptonized tail, with bolometric luminosity
of $\sim 5$\% of that of the disk, with photon index of $\Gamma=2.2$
and electron temperature fixed at 100~keV (i.e. outside the bandpass
of either instrument). The seed photons are assumed to be a blackbody
at the maximum inner disk temperature, set by noting that disk fits to
the peak spectrum of XTE J1817-330 give $kT_{\rm disk}\sim 0.9$~keV
(GDP08). We use this as our fiducial seed temperature for $\log
l=-0.6$, and then change it by a factor $(10^{0.1})^{1/4}=1.059$ for
each 0.1 dex change in $\log l$. We also add 1\% systematic
uncertainties to our simulated PCA spectra as is usual with real data.

These simulated spectra are then fit with a very simple multicolor
disk blackbody model, {\tt diskbb} \citep{mit84}. This assumes that
the disk temperature $T(r)\propto r^{-3/4}$, i.e. has continuous
stress at the inner boundary. The emitting area can still be derived
from the data by applying a correction factor for this \citep{kub98}
as the overall shape of this spectrum is very similar to that produced
by a stress-free inner boundary condition \citep{gme01}.  We also
include a Comptonized tail produced from (blackbody) seed photons tied
to the inner disk temperature and absorption fixed at Galactic values.
The disk so dominates the CCD bandpass that the Compton tail is poorly
constrained, and allowing the spectral index to be free gives best fit
parameters, for which the Comptonized tail is very steep.  This is due
to the {\tt bhspec} disk shape being subtly different to a simple sum
of blackbodies, firstly as relativistic effects smear out the Wien
tail and secondly as the color-temperature correction is not constant
as a function of radius. These combine to give a broader disk spectrum
than predicted by {\tt diskbb} so the low temperature Comptonization
is required to extend the emission to higher energies. Instead, we
make the assumption that simultaneous high energy data exist to
constrain the spectral index (such as from an {\it RXTE} monitoring
campaign: e.g. GDP08), hence we fix this at the input value of 2.2 in
the CCD fits, but allow it to be free in the proportional counter
models.

Fig. \ref{f:lt} shows the resulting $L-T$ diagram from fitting the CCD
and proportional counter simulations for the two new stress
prescriptions considered here (green: beta disk with $\alpha=0.1$,
blue: mean disk with $\alpha=0.1$), together with the two previous
alpha disk models for comparison (black: $\alpha=0.01$ and red,
$\alpha=0.1$). All the stress prescriptions which yield large surface
density ($\alpha=0.01$, mean and beta disks) show very similar behavior.
Only the alpha disk with $\alpha=0.1$ gives significantly higher
temperatures at high luminosities.  However, at low luminosities, all
the stress prescriptions give a fairly good $T^4$ relation, showing
clearly that the \cite{mfr00} result of a substantial increase in
color temperature at low luminosities is indeed an artifact, and is
{\em not} representative of a gas pressure dominated beta disk
\citep[c.f. the discussion in][ where the Merloni et al. result was
used to argue against a beta disk model]{gad04}.

\begin{figure*}
\begin{center}
\hbox{
\includegraphics[width=0.5\textwidth]{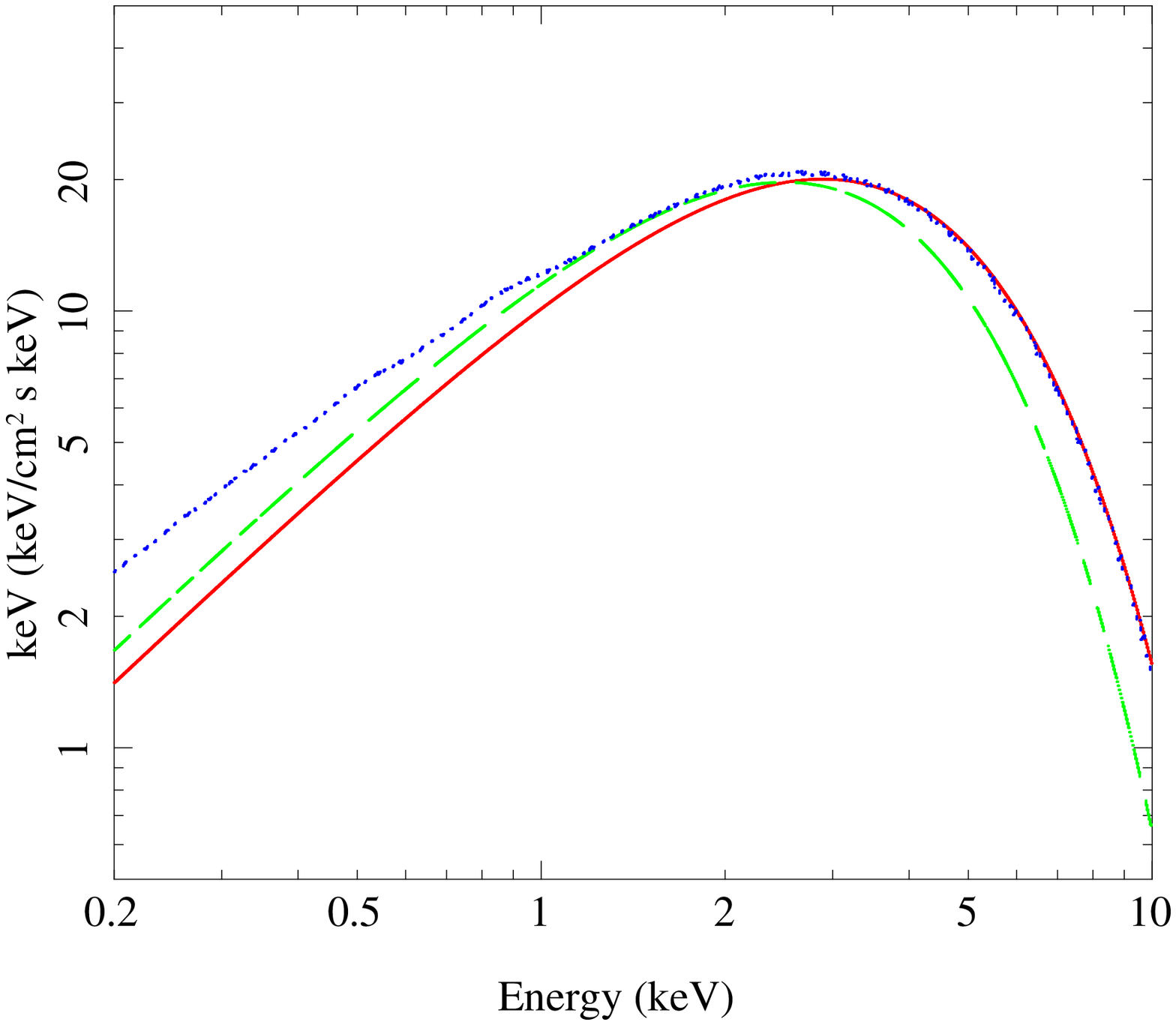}
\includegraphics[width=0.5\textwidth]{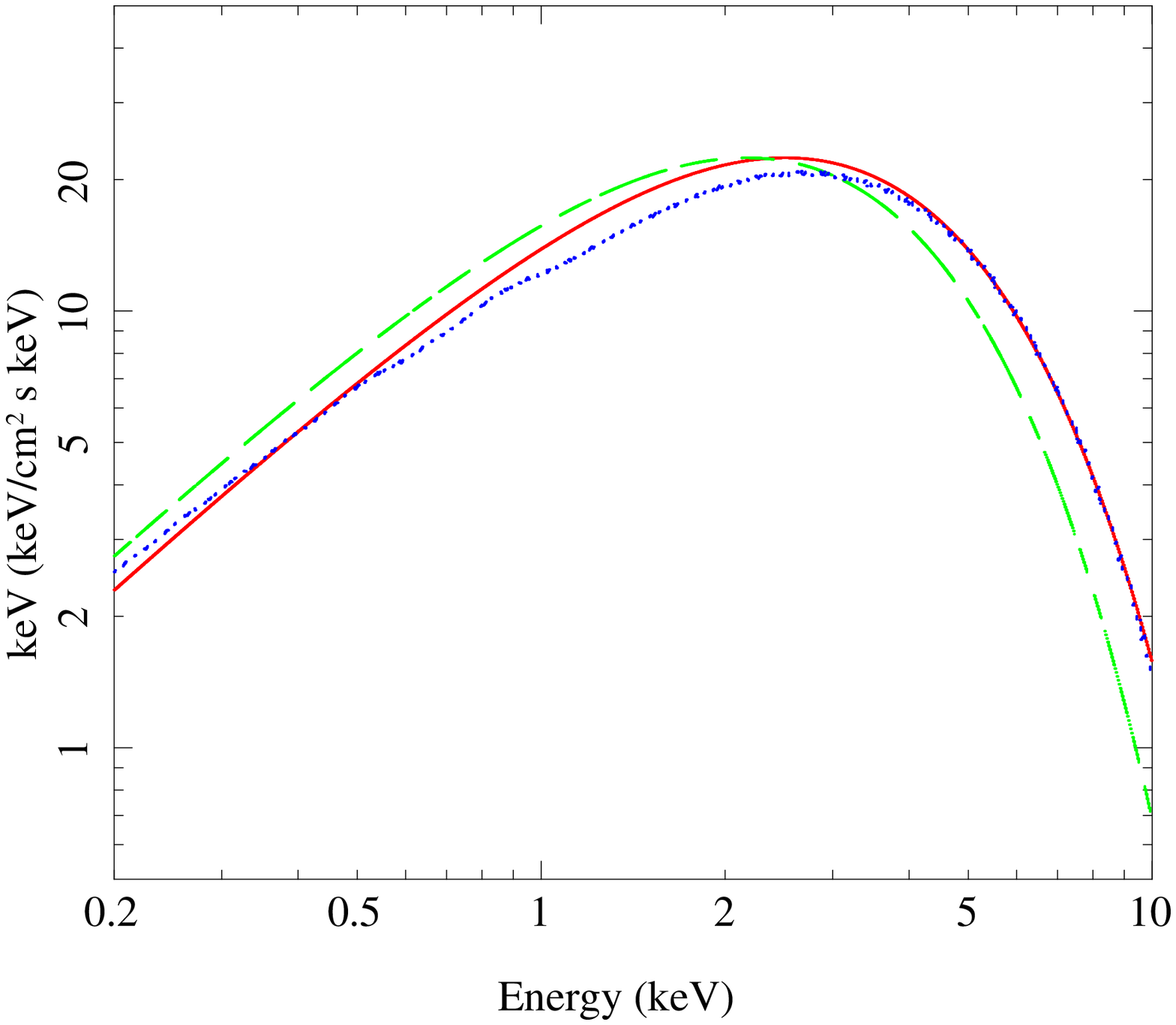}
}
\end{center}
\caption{a) The best-fit {\tt diskbb} models to simulated {\tt bhspec}
disk spectra with $a_*=0.5$ and $\log l=-0.3$ for the proportional
counter (red, solid curve) and CCD (green, dashed curve)
responses. For comparison, we also plot actual {\tt bhspec} model
(blue, dotted curve) used to generate the simulated spectrum. The {\tt
diskbb} spectrum is narrower than {\tt bhspec} so the best fit models
adjust as the bandpass changes. The CCD gives a lower temperature by a
factor 1.15, and the discrepancy at 5-10~keV is taken up by the
freedom in fitting the Comptonized tail.  b) A similar comparison of
{\tt kerrbb} models with the same {\tt bhspec} model plotted in panel
a.  However, in this case the {\tt kerrbb} spectra are {\em not} best
fit models. Instead, they have spin and mass accretion rates set at the
intrinsic parameters for the {\tt bhspec} model, but with 
color-temperature corrections chosen to have the same ratio (1.15, see text
for further details) as the best-fit {\tt diskbb} temperatures.  It is
plain that the color-temperature corrections derived from the {\tt
diskbb} fits are not appropriate for {\tt kerrbb} because the models
have different spectral shapes. \label{f:spec} }
\end{figure*}

The luminosity in these plots is derived as in the $L-T$ plots of GD04
and DGK07 by correcting the disk flux and temperature for inclination
angle and relativistic effects.  The general relativistic corrections
are taken from \cite{zcc97} for a Schwarzschild black hole as this is
closer to an $a_*=0.5$ disk than the alternative tabulation for
$a_*=0.998$.  However, these have very little effect at $60^\circ$, as
this is where the Doppler blueshift and gravitational redshifts
approximately cancel, thus the flux to luminosity conversion is very
close to a simple disk area correction of $L=2\pi D^2 F/\cos i=4\pi
D^2 F$ for $\cos i =0.5$, and the temperature correction is
negligible. However, the recovered flux is less than the input value
(e.g. for the highest luminosity points made from $\log l=-0.2$, while
the dense disks give $-0.26$) as limb darkening is present in
the {\tt bhspec} radiation transfer.

The solid lines show the predicted $L-T$ relation for a constant color
temperature of 1.6, 1.8, 2.0 and 2.2 again as in GD04, but scaling 
the disk area to that expected from an $a_*=0.5$ black hole. Clearly
the CCD data are very close to a constant value of $f_{\rm col}=1.7$,
while the proportional counter bandpass gives a small change in 
$f_{\rm col}$ from 1.8-2.0 for the dense disks, and 1.8-2.2 for the
alpha disk with $\alpha=0.1$. 

The top panels of Fig. \ref{f:l-1} show the effective optical depth of
the disk ($\tau_{\rm eff} \equiv \sqrt{ \tau_{\rm abs}(\tau_{\rm
es}+\tau_{\rm abs})}$) as a function of radius for $\log l=-1$ and
$-0.2$ respectively.  For comparison we also plot the surface density
(middle) and effective temperature (bottom) for the same models.  For
alpha disks the minimum in $\tau_{\rm eff}$ closely corresponds to the
minimum in $\Sigma$. Furthermore, these minima are close to the radius
of maximum $T_{\rm eff}$, so the majority of the flux will be produced
in the regions of the disk which have the lowest opacity and hence the
highest color temperature.  Comparison of the top and middle panels
show that $\tau_{\rm eff}$ is very similar in shape to $\Sigma$.  For
a given alpha, the beta disk always yields a larger $\Sigma$ (and,
therefore, a larger $\tau_{\rm eff}$) than the alpha disk, with the
mean disk midway between them.  Due to the different scalings with $l$
(see Eqs. \ref{eq:siggas} and \ref{eq:sigrad}), the models become more
discrepant as luminosity increases.  The alpha disk with $\alpha=0.1$
(red, solid curve in Fig. \ref{f:l-1}) always has the lowest effective
optical depth, and so gives the largest color-temperature correction.
However, the effect of this is small at low luminosities
(Fig. \ref{f:lt}) as there is still substantial absorption opacity
(Fig. \ref{f:l-1}a) and the heating mainly occurs at high effective
optical depth. Thus, the photosphere acts simply as an atmosphere. By
contrast, at the highest luminosities, the alpha disk with
$\alpha=0.1$ approaches $\tau_{\rm eff}=1$. All of the flux is
dissipated above the effective photosphere itself, and the
color-temperature correction becomes much larger.

We repeat the spectral fitting, replacing the simple {\tt diskbb}
model by the more physical {\tt kerrbb} spectrum \citep{li05}.  Like
{\tt diskbb}, {\tt kerrbb} is a multicolor blackbody model, but it
includes all of the relativistic effects modeled in {\tt bhspec}
i.e. all the special and general relativistic smearing of the
intrinsic disk emission due to rapid rotation in strong gravity. These
broaden the spectrum so that it is not so sharply peaked as predicted
in the {\tt diskbb} models, but unlike {\tt bhspec} it assumes that
the intrinsic emission from each radius is just a color-temperature
corrected blackbody, and that this color correction factor is the same
at all radii.  Therefore, the derived $f_{\rm col}$ will arise solely
from differences in the treatment of the surface emission \citep[see
e.g.][]{sha06}.  We fix the mass, distance, inclination and spin, and
fit for the mass accretion rate and color-temperature correction
assuming no stress on the inner boundary and no returning radiation,
but including limb darkening.  Fig. \ref{f:fcol1} shows these for the
CCD (left panel) and proportional counter (right panel) data
respectively. Again the three dense disk prescriptions all show
similar results, with the alpha disk with $\alpha=0.1$ giving higher
color-temperature correction at high luminosity, as before. However,
the values of the derived color-temperature corrections are {\em
different} to those derived from the {\tt diskbb} approach. The
color-temperature correction for {\tt kerrbb} does {\em not} stay
constant in the CCD bandpass, apparently in conflict with the observed
$L\propto T^4$ relation from {\tt diskbb}. It increases from 1.6--1.9,
which at least goes through the $f_{\rm col}=1.7$ constant value seen
with {\tt diskbb}. By contrast, the proportional counter bandpass
gives values of the {\tt kerrbb} color-temperature correction which
are always lower than those seen from {\tt diskbb}, as well as
spanning a wider range, from 1.35--1.8

\begin{figure*}
\hbox{
\includegraphics[width=0.95\textwidth]{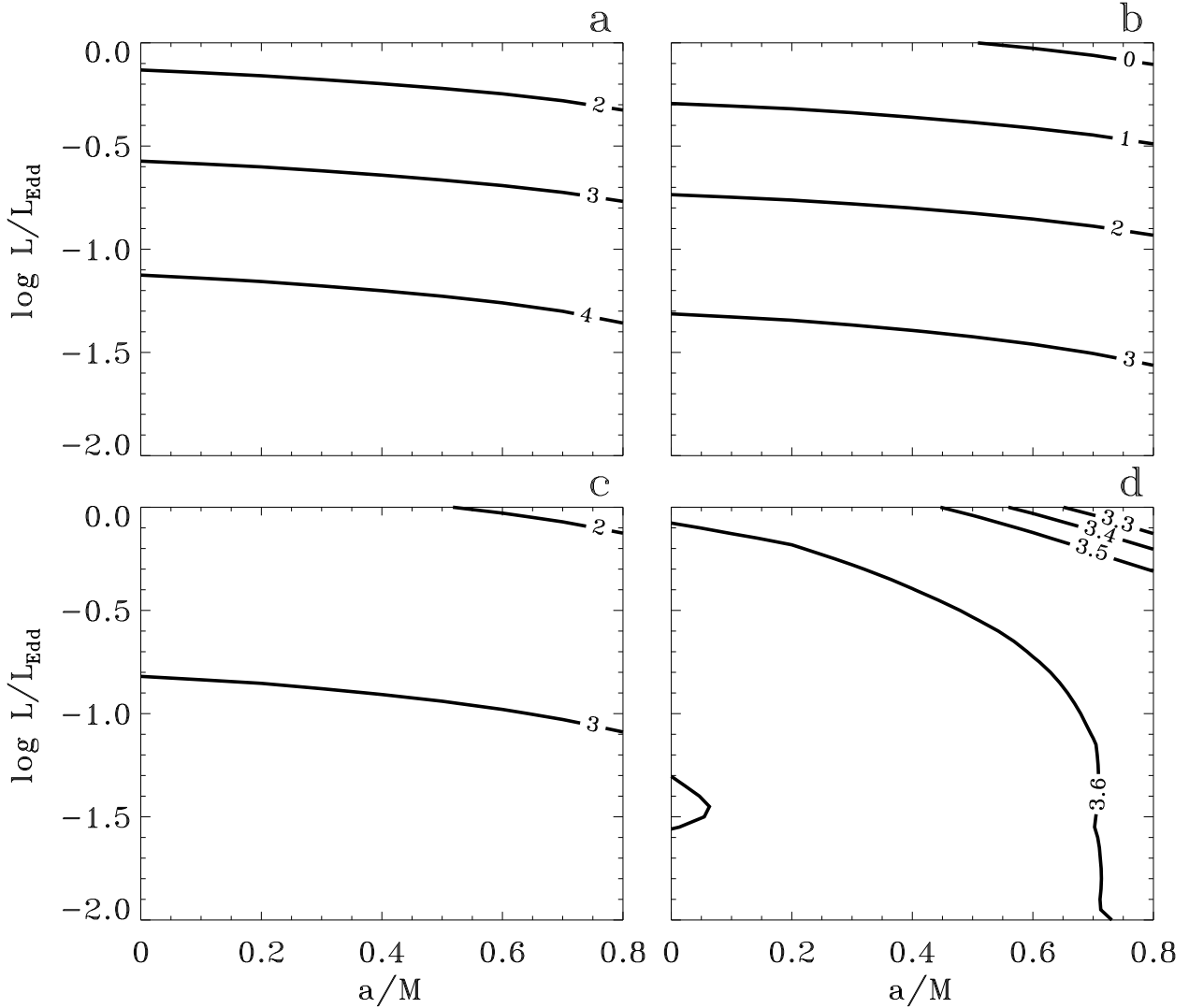}
}

\caption{Effective optical depth at flux maximum as a function of spin
and luminosity. The contour labels are values of $\log \tau_{\rm
eff}$.  The panels correspond to alpha disks with $\alpha=0.01$ (a)
and $\alpha=0.1$ (b), mean disks with $\alpha=0.1$ (c), and beta disks
with $\alpha=0.1$ (d). Note the change in scale of the contours for
panel d. There is less than one order of magnitude change in effective
optical depth across the whole range, so we decrease the spacing to
0.1 dex as opposed to 1 dex for the other panels.\label{f:taue}}
\end{figure*}

The answer to these issues lies in the detailed shape of the
spectrum. Fig. \ref{f:spec}a shows the initial {\tt bhspec} disk
(blue, dotted line) and its best fit {\tt diskbb} spectra to the
proportional counter (red, solid line) and CCD (green, dashed line)
bandpasses for one of the dense disk prescriptions ($\alpha=0.01$) for
an input luminosity of $\log l=-0.3$.  The higher energy bandpass of
the proportional counter weights the fit to higher temperatures/lower
normalization than those for the CCD, and the {\tt diskbb} model is a
good fit to the {\tt bhspec} above the 3~keV lower limit of the
proportional counter bandpass.  However, the best fit to the CCD data
does not look so compelling.  It is clear that the {\tt diskbb}
spectrum is narrower than the {\tt bhspec} disk emission, and the
'best fit' matches this only in the 1-3~keV range where the
signal-to-noise is maximum. At higher energies, the Compton tail
provides additional freedom to match the data.

\begin{figure*}
\hbox{
\includegraphics[width=0.95\textwidth]{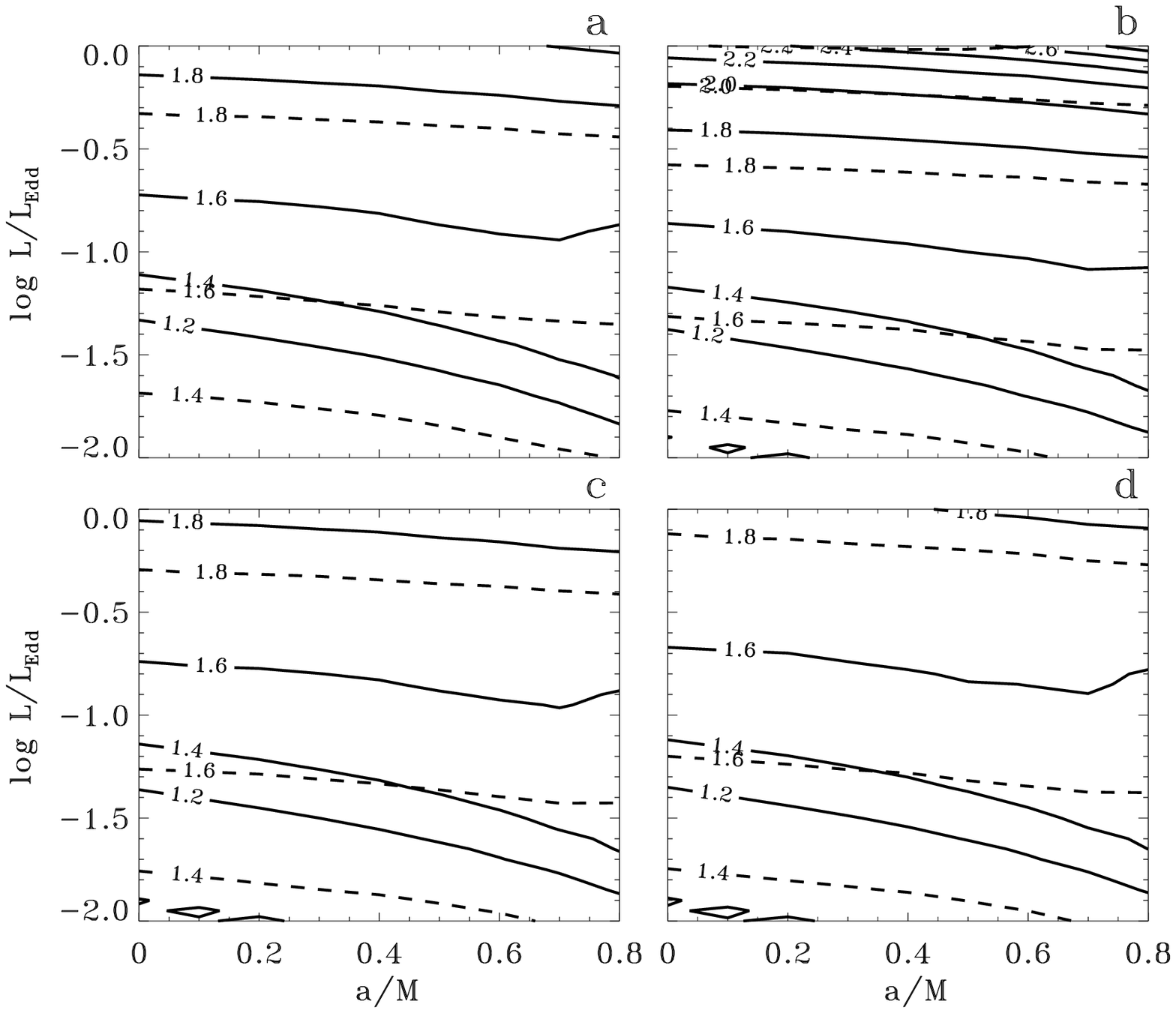}
}

\caption{Color correction at flux maximum for {\tt kerrbb} 
as a function of spin and
luminosity.  The solid and dashed contours are computed using the {\it RXTE}
PCA and {\it Swift} XRT responses, respectively.  See text for more details.
The panels correspond to different stress prescriptions as described 
in Fig. \ref{f:taue}. \label{f:fcol}}
\end{figure*}

Fig. \ref{f:spec}b shows the {\tt kerrbb} models with mass accretion
rate chosen to give $\log l=-0.3$ for a black hole with $a_*=0.5$
(i.e. $8.537\times 10^{18}$~g/s), with color-temperature correction of
1.7 for the proportional counter (red).  The ratio of best fit
temperatures from the {\tt diskbb} models above implies that the
color-temperature correction should be a factor 1.15 smaller in the
CCD compared to the proportional counter i.e.  giving $f_{col}=1.47$
for the CCD (blue).  However, the {\tt kerrbb} spectrum is much
broader than {\tt diskbb} as it includes the relativistic smearing of
the continuum, and the proportional counter model is actually a fairly
good description of the {\tt bhspec} disk shape apart from around the
peak at $\sim 1$~keV. This is because {\tt bhspec} includes the
photoelectric edge features in the atmosphere, calculating the
radiative transfer through the ion populations rather than simply
assuming a diluted blackbody.  These atomic features, from partially
ionized oxygen/silicon/iron L shell, depress the continuum around
1~keV from that predicted by the {\tt kerrbb} models.

Thus the {\tt kerrbb} model appropriate for the proportional counter
data gives a good match below $\sim 0.6$~keV but {\em overestimates}
the flux at the peak. Decreasing the color-temperature correction
simply increases the disparity between the {\tt kerrbb} and {\tt
bhspec} disk spectra in the CCD bandpass, especially in the crucial
1--3~keV region where the signal-to-noise is high. Thus the
appropriate {\tt kerrbb} model would have slightly {\em higher}
color-temperature correction for the CCD than for the proportional
counter, as observed (see Figs. 3a and b).  These differences in
spectral shape mean that color-temperature corrections derived from
{\tt diskbb} cannot be simply applied to {\tt kerrbb} (and vice
versa).

\section{Color-temperature correction and effective optical depth for all spin}
\label{fcol_all}

We use the insights derived above from the particular case of
$a_*=0.5$ to extend our analysis of $\tau_{\rm eff}$ and $f_{\rm col}$
to a larger range of black hole spins and luminosities.  We are
primarily interested the role $\tau_{\rm eff}$ plays in determining
$f_{\rm col}$.  Since our fitting procedure is most sensitive to the
hottest annuli, we will focus only on $\tau_{\rm eff}$ at the radius
where $T_{\rm eff}$ is maximum.  For each $a_*$ and $l$ we calculate
the radius of maximum $T_{\rm eff}$, and evaluate $\tau_{\rm eff}$ at
the midplane for the four stress prescriptions considered in
\S\ref{fcol_0.5}.  Fig. \ref{f:taue} shows contour plots of the
effective optical depth at the radius of maximum flux
(c.f. Fig. \ref{f:l-1}) as a function of spin and luminosity for four
different stress prescriptions. The calculations span a range of $a_*$
from 0 to 0.8 and $\log l$ from -2 to 0. (We consider higher spins
below.)

For comparison, we also fit the {\tt kerrbb} model to simulated {\tt
bhspec} spectra using the PCA and XRT responses.  The procedure we
follow is identical to that described in \S\ref{fcol_0.5}, but we now
calculate $f_{\rm col}$ over the same range of $a_*$ and $l$ as in
Fig. \ref{f:taue}. Fig. \ref{f:fcol} shows the resulting
color-temperature corrections for {\tt kerrbb} fit over a proportional
counter (PCA, solid) and CCD (XRT, dashed) bandpass.  (We stress again
that the different shape of {\tt kerrbb} to that of {\tt diskbb} means
that these are {\em not} applicable to {\tt diskbb} fits.)  Except for
the highest luminosities of the alpha disk model with $\alpha=0.1$,
the CCD response always yields a higher $f_{\rm col}$ than the
proportional counter.

A key issue which determines the spectrum is the fraction of energy
dissipated in the region where thermalization will be incomplete
i.e. above $\tau_{\rm eff}\sim 3$.  This crucial translucent region of
the disk is simply an atmosphere (negligible dissipation) for very
optically thick disks, as the heating occurs at much greater
depths. However, when the whole disk has $\tau_{\rm eff}<30$ then a
substantial fraction of the energy can be emitted in the translucent
region, leading to a large increase in $f_{\rm col}$.
Fig. \ref{f:taue} shows that the alpha disk with $\alpha=0.1$ (panel
b) is the only prescription for which the majority of the flux can be
produced in the region with $\tau_{\rm eff}<3$ (i.e. $\log \tau_{\rm
eff}=0.5$). This happens only for luminosities near Eddington and is
only weakly dependent on spin. A comparison with Fig. \ref{f:fcol}
shows that these models are also the ones with highest $f_{\rm col}$.
The other three stress prescriptions (alpha disks with $\alpha=0.01$,
beta and mean disks with $\alpha=0.1$) tend to give more modest
$f_{\rm col}$ at these $l$.  This comparison also identifies
$\tau_{\rm eff} \simeq 30$ as the point at which dissipation in the
atmosphere becomes important. At lower luminosities ($\log l \lesssim
-0.5$) $\tau_{\rm eff}$ is larger and $f_{\rm col}$ is rather
insensitive to the stress prescription.

\begin{figure*}
\hbox{
\includegraphics[width=0.95\textwidth]{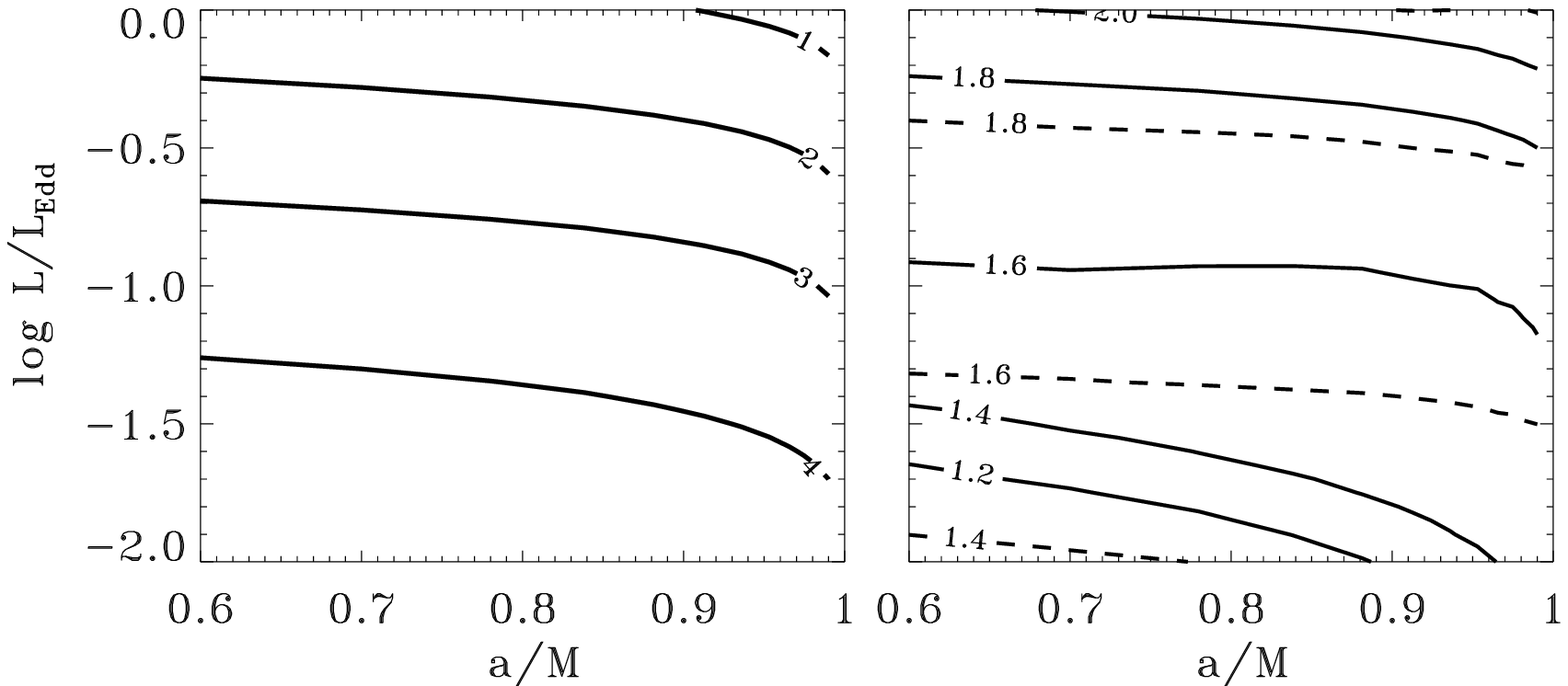}
}

\caption{As in Fig. \ref{f:taue} but extended to high spin 
for an alpha disk with $\alpha=0.01$.  The left panel gives effective
optical depth at flux maximum, while the right panel gives the 
values of $f_{\rm col}$ for {\tt kerrbb} computed using the
{\it RXTE} PCA (solid) and {\it Swift} XRT (dashed) responses, respectively.
\label{f:a0.99}}
\end{figure*}

For these low-to-moderate spins ($a_* \le 0.8$) the variations of
$f_{\rm col}$ and $\tau_{\rm eff}$ have a much stronger dependence on
$l$ than on $a_*$.  These spins are probably the ones most appropriate
for the majority of BHB which have low mass companions, as there is
not enough mass in the companion for the accreted material to
significantly change the angular momentum of the black hole
\citep{kak99}. Thus they fairly accurately reflect the birth spin
distribution from massive star collapse, for which the best estimates
give spin $a_* \lesssim 0.75-0.9$ \citep{gsm04}. However, in the rarer
high mass X-ray binaries \citep[which probably also form the ULX
population e.g.][]{kin08} there is more mass in the companion star, so
accretion may change the mass and spin of the black hole if the
lifetime is long enough. Thus we also investigate higher spins, but
these are computationally problematic for alpha disks with
$\alpha=0.1$ as they become very effectively optically thin for $a_*
\sim 1$ and $l \sim 1$. Since this stress prescription is ruled out by
both spectral and variability behavior for the BHB, we take only alpha
disks with $\alpha=0.01$ as a guide to the behavior of the other dense
disks (mean and beta) as a function of spin for $0.8 < a_* \le
0.99$. Contour plots for $f_{\rm col}$ and $\tau_{\rm eff}$ are shown
in Fig. \ref{f:a0.99}. The contours show a slightly stronger
dependence on $a_*$ at these higher values, with $\tau_{\rm eff}$
decreasing and $f_{\rm col}$ increasing as $a_*$ approaches
unity. Nonetheless, the overall effect is still relatively small
compared to the much stronger dependence on $l$, and the conclusions
drawn from Fig. \ref{f:lt} generalize quite broadly to other spins.
Thus this predicts that any of the dense disk stress prescriptions
will give disk spectra at high spin with $f_{\rm col}$ increasing
slightly with $l$ in a way which is not significantly different to
that for low spin.

Thus extreme spins should simply manifest themselves as higher
temperature disks at a given luminosity, giving a clear signature of a
maximal Kerr black hole.  However, none of the high mass X-ray
binaries (e.g. Cyg X-1) show any signs of this, so we
conclude that accretion has not yet had time to significantly spin up
the black holes in these systems. The ULXs are also potentially high
spin objects \citep{ebi03,hak08}, but these have 
spectra which are often more complex than a simple disk model \citep[see,
however,][]{hak08}, in which case they do not give a
straightforward diagnostic of black hole mass and spin \citep{dak06}.

\section{Discussion}
\label{discus}

\subsection{Stress as a function of radius}

The observation that stable disk spectra are seen spanning $l\sim
0.05-0.5$ shows that disks are not subject to the limit cycles
predicted by the radiation pressure instability in this range. This
rules out an alpha type stress, and instead {\em requires} that the
surface density increases (or remains constant in the limit of
marginal stability) as a function of $l$ at each radius, and this in
itself implies that the accretion disk should remain effectively
optically thick at high luminosities.  Therefore the color-temperature
corrections will be relatively constant, so {\em predicting} the
approximate $L\propto T^4$ relation for the disk dominated
spectra. The observation that most binaries show an approximate
$L\propto T^4$ is strong confirmation of this conclusion.

Sadly, this also rules out the radiation pressure instability as being
the physical mechanism for any observed behavior for $l<0.5$. This is
otherwise an attractive possibility for the origin of the very high
state, the other spectral state seen in binaries at high mass
accretion rates. Although this does not show the predicted limit
cycles, there is the possibility that the non-linear outcome of the
instability could instead lead to strong Comptonization of the disk
\citep[e.g.][]{rac00b,kam04}. Nonetheless, the arguments above
preclude an alpha type stress, at least below $l\sim 0.5$, yet the
very high state is indeed seen in the same range in luminosity as the
disk dominated state \citep[e.g.][]{ram06}.

At higher luminosities \citep[up to $l\sim 3$,][]{dwg04}, there is a
limit cycle observed in GRS1915+104 which could be the remnant of this
radiation pressure instability from a marginally stable stress such as
predicted with the mean disk \citep{hkm91,man06}.  However, this limit
cycle behavior could alternatively be from other instabilities to do
with the super Eddington flows. This is an important distinction, as
the classic radiation pressure instability is not scale invariant with
mass, but is triggered at $l\propto m^{-1/8}$. Thus even if the disk
is stable in stellar mass black holes up to $l\sim 0.5$, it can be
unstable at $l\gtrsim 0.09$ for a $10^7 \msun$ AGN. In this case the
instability could play a role in producing the puzzling, 'soft excess'
seen in high mass accretion rate AGN, predominantly Narrow Line
Seyfert 1's \citep{rac00b,fab02}. However, there are multiple
similarities between the very high state and Narrow Line Seyfert 1
spectra to make it more likely that there is a similar explanation for
both types of object. 

Similarly, those properties of super Eddington flows which likewise
depend on the alpha stress are also probably not a realistic
description of these disks.  Radiation trapping is a generic feature
of any of the stress prescriptions \citep[e.g.][]{hkm91}, resulting in
optically thick advection of energy especially in the disk midplane
\citep{wmm01}. However, the denser disks which result from the
alternative stress prescriptions are less likely to become effectively
optically thin to the escaping radiation, so are unlikely to show the
very large color-temperature correction which can arise from
overheating of alpha disks \citep[e.g.][]{bel98,kaw03}. We caution
that fitting such models to high mass accretion rate spectra from ULX
and NLS1's may not be appropriate.

\subsection{Caveats}

So what then are the caveats to using disk spectra as an estimator for
black hole spin? The first is that the spectra {\em must} be disk
dominated.  The disk radius can be under -- or overestimated when a
substantial fraction of the dissipation goes instead into a corona,
either in the very high \citep{kad04,dak06} or low/hard state (GDP08).
This may be the origin of the discrepancy in spin determination by disk
spectral fitting in GRS 1915+104 \citep[compare][]{mid06,mcc06}. 
The spectra of GRS 1915+104 may well be more
complex than the disk dominated spectra seen from BHB due to its
higher luminosity \citep{dwg04}, perhaps 
powering strong winds (see below). It seems premature to apply the
uncertainties from this one pathological object to other sub-Eddington
BHB \citep[e.g.][]{raf08}.

Secondly, the value of the color-temperature correction is robust for
all of the dense disk stress prescriptions, but varies with bandpass
and also varies with the detailed disk model used to fit the
spectra. We caution that differences in spectral shape between the
models means that the same color-temperature correction factors {\em
cannot} be simply applied to {\tt diskbb} and {\tt kerrbb}
fits. Ironically, for both the generic proportional counter and CCD
bandpasses, the {\tt diskbb} fits give a simpler (approximately
constant color temperature) representation of the data.

Thirdly, the structure of the disk as $l\to 1$ is not well modeled by
this code.  We neglect optically thick advection, and changes in the
scale height of the disk with radius which may lead to self shielding
for highly inclined objects \citep[e.g.][]{wtf05}, but these are
unlikely to have much effect for $l<0.5$, where the majority of BHB
data are taken.

However, disks at high luminosities are likely to also power
substantial winds.  There are clear signs of an equatorial disk wind
from observations of highly ionized, blueshifted, Fe K$\alpha$
absorption lines. These are ubiquitously seen in CCD data from
high/soft state BHB (i.e. $l>0.1$) with inclination angles $\ge
60^\circ$ (e.g. DGK07). This material is probably a thermally driven
disk wind, where irradiation of the outer disk layers produces a
Compton heated skin whose high temperature allows gas particles to
have enough energy to escape at large radii \citep{bms83,woo96},
though this may be (substantially?)  enhanced by magnetic fields
\citep{mil06}. Such winds almost certainly become more powerful with
$l$, especially as $l\to 1$ and radiation driving enhances them still
further \citep{pak02}. Counterintuitively, these stronger winds may
become less observable as the stronger irradiation means that they
become more ionized, losing even the He- and H-like Fe
\citep{woo96}. Thus there may be winds which have substantial column,
but are unobservable as they are completely ionized.  Electron
scattering has the effect of suppressing the observed luminosity by a
factor $e^{-\tau}$ for lines of sight intersecting the wind, while the
scattered flux $(1-e^{-\tau})$ can enhance the luminosity along the
disk axis by a factor $1+(1-e^{- \tau})\Omega/4\pi$ where $\Omega$ is
the solid angle subtended by the wind. This may be the origin of the
occasionally observed `bends' in the $L-T$ diagram, where the data
deviate away from the generally observed slow increase in $f_{\rm
col}$ as a function of luminosity e.g. in GRO J1655-40 there is a much
stronger increase in $f_{\rm col}$ with $l$ than predicted by the
dense disk models (e.g. Fig 1b), while in LMC X--3 there is a slight
{\em decrease} \citep[see e.g.][]{ddb06}.  While the inclination of
LMC X-3 is not well known, GRO J1655-40 is plainly at very high
inclination, and has clear wind features detected in multiple
observations \citep{mil06,sal07}.

Thus as well as avoiding spectra with strong tails, and super
Eddington sources, we also caution against any serious attempt to
derive spin from `bent' $L-T$ diagrams, where winds may be important.

Our results also hinge on the fact that the dissipation above the
effective photosphere is small when the surface density is
sufficiently large, so our assumptions about the dissipation profile
with height may be an important caveat.  Previous work
\citep{dav05,bla06} has demonstrated that dissipation profiles
motivated by radiative, stratified shearing box simulations
\citep{tur04,hks06} do not put enough dissipation near the surface to
significantly alter the spectrum, although a more exhaustive range of
parameter space needs to be explored. However, the alternative stress
prescriptions discussed here are partially motivated by theoretical
arguments that limit the magnetic stresses due to transport of buoyant
magnetic field \citep[see e.g.][]{sar84,mer03}.  If transport of
magnetic field is significantly larger than currently seen in
simulations, dissipation would be concentrated much nearer the surface
than our models assume \citep{mer03}. This would give significant
dissipation above the photosphere, and hence lead to a marked increase
in color-temperature correction compared to those calculated here for
the dense disks.  As long as this is not a strong function of
luminosity then it could still fit the data, but would lead to the
spin being overestimated. Since the spins are generally
low-to-moderate, there is not much scope for the real stresses to
produce a significant fraction of the dissipation within the
photosphere.

Similarly, the data also preclude a large amount of dissipation at the
last stable orbit. If this were to thermalize it would lead to higher
temperature emission, so current models would overestimate the spin,
yet these already favor low spins. Alternatively, if the dissipation
did not thermalize, it would instead produce a substantial
non--thermal component in the spectrum, yet the spectra are dominated
by the disk emission in the high/soft state.  Thus there is only
limited scope for stresses at the last stable orbit in disk dominated
spectra, in contrast to the results from non-radiative, relativistic
magnetohydrodynamic simulations \citep[e.g.][]{bhk08}. It may be that
the character of the flow qualitatively changes when radiative cooling
is included in the simulations, as the small scale height fields
sustained by a thin (radiatively cooled) disk may lead to much less
stress at the last stable orbit than the large scale fields which can
be generated in the large scale height, non-radiative flows
\citep{aap03,snm08}.

This potential difference in dissipation between large scale height
flows and thin disks may actually be observed in the behavior of the
transition between the low/hard state, where the spectrum is produced
by Compton scattering in a hot, optically flow, and the disk dominated
state where the emission is quasi--thermal. The low/hard state is
probably produced by a large scale height flow, which has an
efficiency which is at least a factor 3 lower than that of a
(stress--free inner boundary condition) thin disk down to the same
radius \citep[e.g.][]{nay95}.  This predicts that the luminosity at
the transition should jump by at least a factor 3, yet the data show a
smooth transition of less than a factor $\sim 2$ \citep[e.g. the
compilation black hole binaries, especially Cyg X-1,
in][]{dag04}. This can be explained if the large scale height, less
efficient flow in the low/hard state has continuous dissipation across
the last stable orbit due to large scale height magnetic fields. In
effect this allows the flow to extend down to smaller radii, so it
taps a larger fraction of the gravitational potential energy, hence
somewhat compensating for its lower radiative efficiency compared to
that of the thin disk (Done et al. 2008, in preparation).

\section{Conclusions}
\label{conc}

The detailed nature of angular momentum transport remains a
significant uncertainty in our understanding of accretion flows and
their emission. It is not yet entirely certain that the MRI, which is
the best candidate for the source of the turbulent stress, truly does
saturate at a level high enough to satisfy observational constraints
\citep{pcp07,fap07,kpl07}.  Even if this question is resolved in favor
of the MRI, a detailed understanding of the stress and resulting disk
structure will likely require fully-relativistic, global simulations
with realistic thermodynamics.  Since such simulations require
significant advances in computing power, we are forced rely on much
simpler parameterizations of the stress until these become available.
Even though these prescriptions are ad hoc, they may still capture
important aspects of the physics in real astrophysical flows and
provide useful constraints by direct comparison with observations.

With this motivation, we have computed the accretion disk spectra
predicted by several different stress prescriptions using the most
complete spectral code currently available. This includes both non-LTE
ion populations, with radiative transfer in the disk and full general
relativistic ray tracing to propagate this flux to the observer. These
models give results which are very close to the observed behavior of
the spectra of black hole binaries in the thermal dominant (high/soft)
state from $L/L_{Edd}\sim 0.06\to 0.6$.  The majority of this data
comes from proportional counters, and generally shows that the
temperature changes slightly more rapidly with luminosity than
expected for a disk of constant inner radius, and constant
color-temperature correction (e.g. the compilation of DG04 and DGK07).
This is as predicted by all the high surface density disk models
presented here, and should hold generally for all stress prescriptions
where less than 10 per cent of the energy is dissipated above the
effective photosphere.  The surface layers then simply act like a
passive atmosphere, with properties set by the effective
temperature. The increase in temperature with luminosity gives rise to
a small increase in color-temperature correction and the resulting
spectra are remarkably similar irrespective of the detailed form of
the stress.

Such model fitting can also place interesting constraints on the
angular momentum transport. The classic alpha disk with $\alpha=0.1$
becomes effectively optically thin at the highest luminosities. The
resulting increase in color temperature with luminosity is much more
rapid than observed from existing data, ruling out such stress
prescriptions. The standard alpha disk is also unstable to the
thermal-viscous radiation pressure instability which predicts limit
cycle behavior which is not observed, again showing that the disk is
denser than such models predict.  However, even these models give the
same spectra at low luminosities ($L \lesssim 0.1 L_{\rm Edd}$), where
the surface density is sufficiently large for the disk to remain very
effectively optically thick.  Therefore, where we have a clear view of
the disk, unaffected by a moderately optically thick disk wind, the
disk spectra should provide a relatively robust estimator of the disk
inner radius, and plausibly, the spin of the black hole.

\acknowledgements{We thank K. Beckwith, O. Blaes, M. Gierli{\'n}ski,
I. Hubeny, J. Krolik, and A. Kubota for useful discussions.  CD
acknowledges support from a PPARC senior fellowship, and a Royal
Society conference grant and Omer Blaes who together funded the visit
to Santa Barbara at which this paper was begun.  SD is supported by
NASA grant number PF6-70045, awarded by the Chandra X-ray Center,
which is operated by the Smithsonian Astrophysical Observatory for
NASA under contract NAS8-03060.}

\end{document}